%% file: main.tex
\documentclass[letterpaper, 10 pt, conference]{ieeeconf}  

\IEEEoverridecommandlockouts                              
                                                          
\usepackage{epsfig}
\usepackage{url}
\usepackage{amsmath}
\usepackage{amsfonts}

\usepackage{amsthm}
\usepackage{amssymb}
\usepackage{graphicx}
\usepackage{color}
\usepackage{balance} 
\usepackage{algorithm2e}
\usepackage{algorithmic}
\usepackage{bm}

\newtheorem{prop}{Proposition}

\newtheorem{definition}{Definition}

\let\labelindent\relax

\usepackage{enumitem}
\usepackage{graphicx} 
\graphicspath{{figure/}}
\usepackage{float}
\usepackage[caption=false,font=footnotesize]{subfig}
\usepackage{soul}
\usepackage{color}

\usepackage{algorithm,algorithmic}
\usepackage{cite} 
\newcommand\highlight[1]{{#1}}

\title{\LARGE \bf Lyapunov-Regularized Reinforcement Learning for Power System \\ Transient Stability }

\author{Wenqi Cui and Baosen Zhang
\thanks{Department of Electrical and Computer Engineering, University of Washington Seattle, WA 98195, USA	 \{wenqicui, zhangbao\}@uw.edu}%
\thanks{The authors are supported in part by the National Science Foundation grant ECCS-1930605 and the Washington Clean Energy Institute.}%
}

\begin{document}
\maketitle
\thispagestyle{empty}
\pagestyle{empty}

\begin{abstract}
Transient stability of power systems is becoming increasingly important because of the growing integration of renewable resources. These resources lead to a reduction in mechanical inertia but also provide increased flexibility in frequency responses. Namely, their power electronic interfaces can implement almost arbitrary control laws. To design these controllers, reinforcement learning (RL) has emerged as a powerful method in searching for optimal non-linear control policy parameterized by neural networks. 

A key challenge is to enforce that a learned controller must be stabilizing.  
This paper proposes a Lyapunov regularized RL approach for optimal frequency control for transient stability in lossy networks. Because the lack of an analytical Lyapunov function, we learn a Lyapunov function parameterized by a neural network. The losses are specially designed with respect to the physical power system. The learned neural Lyapunov function is then utilized as a regularization to train the neural network controller by penalizing actions that violate the Lyapunov conditions. Case study shows that introducing the Lyapunov regularization enables the controller to be stabilizing and achieve smaller losses.



\end{abstract}

\section{Introduction}
\input{introduction}

\section{Model and Problem Formulation} \label{sec:model}
\input{model}

\section{Learning a Lyapunov function} \label{section:Lyapunov}
\input{Lyapunov}

\section{Learning Neural Network Controller with Lyapunov regularization} \label{section:Controller}
\input{Controller}

\section{Case Study} \label{sec:simulation}
\input{simulation}

\section{Conclusion} \label{sec:conclusion}
This paper proposes a Lyapunov regularization approach to guide the training of neural network controller for primary frequency response for transient stability. A function paramertized as neural network is learned to overcome the difficulty brought by the non-existence of analytical Laypunov functions for lossy power networks. By integrating the neural Lyapunov function as a regularization term for the training of neural network controller in RL, control actions that violate Lyapunov conditions are penalized. Case studies verify introducing Lyapunov regularization enable the controller to be stabilizing and achieve smaller losses, whereas controllers trained without regularization can fail to stabilize the system. \highlight{ An important future direction for us is to understand the region of attraction better in the context of learning controllers.}

\bibliographystyle{IEEEtran}
\bibliography{Reference}
\appendices
\input{appendix}
\end{document}

%% file: introduction.tex
Transient stability in power systems refers to the ability of a system to converge to an acceptable steady-state after a disturbance~\cite{chiang1989study, arapostathis1982global}. With the increased penetration of renewable energy sources (RES), power systems have reduced inertia and transient stability is becoming increasingly important~\cite{jiang2020dynamic}. Meanwhile, RES are connected to the grid via electronic interfaces and can be controlled freely by inverters to implement almost arbitrary control laws. Instead of linear droop frequency response found in conventional generators, the response of the inverter-based RES can be optimized to improve performance by implementing more flexible control laws~\cite{johnson2013synchronization}.

Transient stability models describe how frequency changes in a system with a large deviation of operating states, and use the full nonlinear AC power flow equations~\cite{chiang1989study}. Two challenges emerge in controller design. Firstly, the problem is over a functional space, which is infinite-dimensional. Secondly, the controllers should be stabilizing, which is a nontrivial constraint to enforce algorithmically for nonlinear systems.

A popular way to address the first challenge is to  parameterize the controllers (e.g., using a neural network) and training them using reinforcement learning (RL)~\cite{chen2021reinforcement}. Abundant algorithms, including Q-learning, deep direct reinforcement learning (DDPG), actor-critic, have been proposed for optimal control (see, e.g.,~\cite{sutton2018reinforcement} and the reference within). References~\cite{yan2018data,chen2020model,duan2019deep,ernst2008reinforcement} apply these algorithms for power system frequency regulation. However, the stabilizing requirement of the controllers is not considered in these works.


The challenge of ensuring controllers are stable is more difficult to address. If a Lyapunov function is available, it can potentially provide analytical constraints on the controller. For lossless power systems, using a well-known energy function~\cite{arapostathis1982global,dobson1989towards}, our previous work in~\cite{cui2020} showed how to impose structural constraints on the neural network controllers such that they are guaranteed to be stabilizing. Unfortunately, for lossy networks, there are no known analytic energy functions~\cite{chiang1989study}. Most transmission lines have non-zero resistances, and distribution systems can have high $r/x$ ratios.

If analytical Lyapunov functions are not available, a natural approach would be to learn a Lyapunov function to facilitate controller design. 
For example, given input/output data and the assumption that the underlying system is stable, reference~\cite{manek2020learning} learns a Lyapunov function jointly with learning the system model to find stable system dynamics.  
The work in~\cite{chang2020neural} uses satisfiability modulo theories solvers to formally verify a function satisfies the Lyapunov conditions. However, it is only currently computational tractable for small systems. \highlight{Reference~\cite{huang2020transient} applies this method to distribution system by aggregating networked microgrids as a single node. }Moreover, the above works focus on verifying a system is stable and do not include controller design.



This paper proposes a Lyapunov regularization approach to guide the training of neural network controller for primary frequency response in lossy power systems. We learn a Lyapunov function parameterized by a neural network. The loss function for training the neural Lyapunov function is designed to satisfy the positive definiteness of its value and the negative definiteness of its Lie derivative. Existing methods in ~\cite{manek2020learning,chang2020neural,huang2020transient} weigh all the states equally in the loss function, but this will cause the sub-optimum of Lyapunov function near the equilibrium since the magnitude of states' time derivative shrink quickly when approaching the equilibrium. Considering that the states near the equilibrium are more important for control, we specially design the loss function such that the area around the equilibrium is emphasized. 




The neural Lyapunov function is utilized as a regularization to train the neural network controller by penalizing actions that violate the Lyapunov conditions. The regularized RL is integrated in the \highlight{recurrent neural network} (RNN) based framework in our previous work to increase its training efficiency~\cite{cui2020}.
Simulation results show that the learned function satisfies the Lyapunov conditions for almost all points in the state space, thus making it a good tool for regularization. Case study shows that introducing the Lyapunov regularization enables the controller to achieve smaller loss. More importantly, a controller designed without regularization can lead to unstable behaviors. All of the code and data described in this paper are publicly available at https://github.com/Wenqi-Cui/Lyapunov-Regularized-RL.  One important future work is to verify whether the learned function satisfies the Lyapunov conditions for all points in a region.



%% file: model.tex
\subsection{Frequency Dynamics}
Let $N$ be the number of buses and $\mathcal{E}$ be the set of transmission lines connecting the buses. The susceptance and conductance of the line $(i,j) \in \mathcal{E}$ are $B_{i j}=B_{ji }$ and $G_{i j}=G_{ji}$, respectively; and 0 if the buses are not connected. We use the Kron reduced model to aggregate load buses into generator buses~\cite{nishikawa2015comparative,Tinu20}. We assume that each bus $i$ has the conventional inertia $M_i$ and the damping from synchronous generator and loads is denoted as $D_i$~\cite{machowski2020power, zhao2016unified}. \highlight{    Denote the generator power and load of bus $i$ as $P_{g,i}$ and $P_{l,i}$, respectively. Then, $P_i=P_{g,i}-P_{l,i}$ represents the net power injection of bus $i$.} \highlight{We assume that the synchronous generation are set to their nominal operating points. Our control comes from the inverter-connected resources such as storage and wind turbines~\cite{muljadi2012understanding}. Without loss of generality, we assume that each bus has an inverter-connected resources (the actuation bounds can be both set to zero if a resource is not present).} 

The angle and frequency \highlight{deviation} of bus $i$ are $\delta_i$ and $\omega_i$, respectively. We assume that the bus voltage magnitudes are 1 p.u. and the reactive power flows are ignored.
The dynamics of the power system is represented by the swing equation~\cite{kundur1994power}
\begin{subequations}\label{eq:Dynamic}
	\begin{align}
	\dot{\delta_i} =&\omega_i \quad,\forall i =1,\cdots, N \label{eq:Dynamic_delta}\\
	\begin{split}
	M_{i}\dot{\omega}_{i} =&P_{i}-D_i \omega_i -u_i(\omega_i)-\sum_{j=1, j\neq i}^{N}B_{i j}\sin(\delta_i-\delta_j)\\
	&-\sum_{j=1, j\neq i}^{N}G_{i j}\cos(\delta_i-\delta_j), \quad\forall i =1,\cdots, N  \label{eq:Dynamic_w}
	\end{split}
	\end{align}
\end{subequations}
where $u_i(\omega_i)$ is the controller that changes active power to provide primary frequency response. Because power systems do not have real-time communication infrastructure, we restrict $u_i$ to be a static feedback controller where only its local frequency measurement $\omega_i$ is available. We envision the control is provided by renewable energy resources such as batteries and solar PV. In the primary frequency regulation timescale from 100ms to few seconds for primary frequency regulation, the main limitation on actuation comes from power injection constraints.

\subsection{Optimization Problem Formulation }
The objective is to minimize the cost on frequency deviations and the control effort. In this paper, we use frequency nadir, which is the infinite norm of $\omega_i(t)$ over the time horizon from 0 to the time  $T$ defined as $||\bm{\omega_i}||_{\infty}=\sup_{0\leq t\leq T} |\omega_i(t)|$~\cite{tabas2019optimal}. We use a quadratic cost for the control actions defined by $||\bm{u_i}||_2^2=\frac{1}{T} \int_{t=0}^{T} (u_i(t))^2dt$~\cite{dorfler2013synchronization,Tinu20}. We aim to find an optimal stabilizing controller $\bm{u}(\cdot)$ by solving \eqref{eq:Continuous_Optimization}.
\begin{subequations}\label{eq:Continuous_Optimization}
	\begin{align}
	\min_{\mathbf{u}}\quad & \sum_{i=1}^{N} ||\bm{\omega_i}||_{\infty}+\gamma||\bm{u_i}||_2^2\label{subeq:Continuous_Optimization_obj}\\
	\mbox{s.t. } \quad &\eqref{eq:Dynamic_delta}-\eqref{eq:Dynamic_w}\label{subeq:Continuous_Optimization_dynamics}\\
	& \underline{u}_i \leq u_i(\omega_i)\leq \overline{u}_i\label{subeq:Continuous_Optimization_bound}\\
	& u_i(\omega_i) \text{ is stabilizing}\label{subeq:Continuous_Optimization_stability}
	\end{align}
\end{subequations}
where $\gamma$ is a tradeoff parameter between cost of frequency deviation and action. The swing equations are in~\eqref{subeq:Continuous_Optimization_dynamics}. The controller are power limited within the upper bound $\overline{u}_i$ and lower bound $\underline{u}_i$   in~\eqref{subeq:Continuous_Optimization_bound}. We impose the condition that the controller should be stabilizing in~\eqref{subeq:Continuous_Optimization_stability}.  Constraints~\eqref{subeq:Continuous_Optimization_dynamics}-\eqref{subeq:Continuous_Optimization_stability} hold for the time $t$ from 0 to $T$. \highlight{ Other objective functions can also be used (e.g., $l_1$ penalty on total frequency deviation and the rate of change of frequency) without changing the framework.}

Problem~\eqref{eq:Continuous_Optimization} is challenging to solve by conventional control techniques and we will use RL to find $\bm{u}(\cdot)$. The key difficulty is to quantify the stability requirement in~\eqref{subeq:Continuous_Optimization_stability}. We mitigate this difficulty by using a Lyapunov function, which provides algebraic conditions for~\eqref{subeq:Continuous_Optimization_stability}. Since a Lyapunov function is not known for lossy systems~\cite{chiang1989study}, we show how one can be learned in the next section.




%% file: Lyapunov.tex
\subsection{Lyapunov Conditions}

From standard system theory, the Lyapunov function need to satisfy conditions on its value and its Lie derivatives~\cite{sastry2013nonlinear}. Let the state space be $\mathcal{D} =\left\{(\delta, \omega)| \delta=(\delta_1,\cdots,\delta_N), \omega=(\omega_1,\cdots,\omega_N)\right \}$. The state transition dynamics~\eqref{eq:Dynamic} is written as $(\dot{\delta},\dot{\omega})=f_u(\delta,\omega)$, where $f_u$ stands for the state transition function with respect to the controller $u$. Using the notation from~\cite{chang2020neural}, we have  

\begin{definition}[Lie Derivatives]
 The Lie derivative of the continuously differentiable scalar function $V: \mathcal{D} \rightarrow \mathbb{R}$ over the vector field $f_{u}$ is defined as
\begin{equation} \label{eq: Lie_derivative}
 \begin{aligned} 
 \nabla_{f_{u}} V(\delta,\omega) &=\sum_{i=1}^{N}\frac{\partial V(\delta,\omega) }{\partial \delta_i} \dot{\delta_i}+\frac{\partial V(\delta,\omega) }{\partial \omega_i}\dot{\omega_i}
\end{aligned}
\end{equation}
\end{definition}
It measures the rate of change of $V$ along the direction of the system dynamics. The next proposition is standard in nonlinear systems. 

\begin{prop}
[Lyapunov function and asymptotic stability]
Consider a controlled system described by (1) with equilibrium at $(\delta^{*},\omega^{*} )$. Suppose there exists a continuously differentiable function $V: \mathcal{D} \rightarrow \mathbb{R}$ that satisfies the following conditions
\begin{subequations}\label{eq:Lyapunov_Condition}
\begin{align}
& V(\delta,\omega)>V(\delta^{*},\omega^{*} )\quad  \forall (\delta,\omega)\in \mathcal{D} \backslash\{(\delta^{*},\omega^{*} )\} \label{eq:Lyapunov_Condition_V}\\
& \nabla_{f_{u}} V(\delta,\omega)<0 \quad \forall (\delta,\omega)\in \mathcal{D} \backslash\{(\delta^{*},\omega^{*} )\}
\label{eq:Lyapunov_Condition_dot_V}\\
& \nabla_{f_{u}} V(\delta^*,\omega^*)=0, \label{eq:Lyapunov_Condition_V0}
\end{align}
\end{subequations}
Then the system is asymptotically stable at the equilibrium.

\end{prop}

In this paper, Lyapunov function is parameterized using neural network with weights $\phi$, and written as $V_{\phi}\left(\delta,\omega\right)$. For differentiability, we use \highlight{Exponential Linear Unit}  (ELU) activation functions. Note that $V_{\phi}\left(\delta,\omega\right)$ is purely a function of the state variable $(\delta,\omega)$, while $\nabla_{f_{u}} V(\delta,\omega)$ will be affected by the controller $u_i$ through the term $\dot{\omega_i}$ in \eqref{eq: Lie_derivative}. Therefore, only $\nabla_{f_{u}} V(\delta,\omega)$ will be utilized to regularize controller once it is learned.


\subsection{Learning the Lyapunov Function}
 
The condition \eqref{eq:Lyapunov_Condition_V} is easy to be satisfied if we explicitly engineer the structure of $V_{\phi}\left(\delta,\omega\right)$.
To name a few, $V_{\phi} (\delta,\omega)$ can be formulated using a convex function achieving the minimum at the equilibrium. Or, given an arbitrary function $g(\delta,\omega)$ and positive scalar $\epsilon$, \eqref{eq:Lyapunov_Condition_V}  can be enforced by taking $V_{\phi} (\delta,\omega)=\left(g(\delta,\omega)-g(\delta^*,\omega^*)\right)^2+\epsilon ||(\delta,\omega)-(\delta^*,\omega^*)||_2$.  However, such parameterization may be too restrictive and make it hard to satisfy~\eqref{eq:Lyapunov_Condition_dot_V}. Therefore, we do not explicitly engineer the structure in satisfying condition \eqref{eq:Lyapunov_Condition_V}. 

In this paper we use loss functions to penalize violations of~\eqref{eq:Lyapunov_Condition_V}-\eqref{eq:Lyapunov_Condition_V0}. Training is implemented in a batch updating style where the number of batch is $H$ and the state of the $h$-th batch is randomly generated  $(\delta^h,\omega^h)\in \mathcal{D}$ for $h=1,\cdots,H$. The losses are designed with respect to the following considerations:

\begin{enumerate}[leftmargin = 10 pt]
    \item \textit{Avoid overfitting when $\dot{\delta}$ and $\dot{\omega}$ are large}
    
    \highlight{To satisfy~\eqref{eq:Lyapunov_Condition_dot_V}, the loss term need to encourage  $\nabla_{f_{u}} V(\delta,\omega)$ to be negative and penalize its positive values.} A loss that weighs all points in the space equally leads $\nabla_{f_{u}} V(\delta,\omega)$ to have very negative values when $\delta$ and $\omega$ are far away from the equilibrium, and may violate~\eqref{eq:Lyapunov_Condition_dot_V} for points close to the equilibrium. This contradicts the premise that the small region around the equilibrium should be stabilizing. Therefore, we design the loss term with $\nabla_{f_{u}} V(\delta,\omega)$ to be 
    \begin{equation}\label{eq:Loss_Lya_1}
    \begin{split}
        l_1{(\phi)}=\frac{1}{H}\sum_{h=1}^H &\operatorname{tanh}\left( \nabla_{f_{u}} V_{\phi}(\delta^h,\omega^h)\right)\\
        &\cdot\operatorname{exp}\left(-\frac{||(\delta^h,\omega^h)-(\delta^*,\omega^*)||_2 }{\mu }\right)
    \end{split}
    \end{equation}
    where the term $\operatorname{tanh}\left( \nabla_{f_{u}} V_{\phi}(\delta^h,\omega^h)\right)$ avoid the overfit of $\nabla_{f_{u}} V(\delta^h,\omega^h)$ to be extremely negative \highlight{We use tanh function to make $l_1{(\phi)}$ to have the same sign with as $\nabla_{f_{u}} V_{\phi}(\delta^h,\omega^h)$. }
    The term $\operatorname{exp}\left(-\frac{||(\delta^h,\omega^h)-(\delta^*,\omega^*)||_2 }{\mu }\right)$ emphasis the importance of $(\delta^h,\omega^h)$ closer to the equilibrium. The hyper-parameter $\mu$ controls rate of decay.

    \item \textit{Penalty term with  $\left(V_{\phi}\left(\delta^*,\omega^*)\right)-V_{\phi}(\delta,\omega)\right )$}
    
    In order to satisfying condition \eqref{eq:Lyapunov_Condition_V}, $V_{\phi}(\delta^h,\omega^h)$ that is smaller than $V_{\phi}\left(\delta^*,\omega^*\right)$ need to be penalized. \highlight{ For points that satisfy $V_{\phi}(\delta^h,\omega^h)>V_{\phi}\left(\delta^*,\omega^*\right)$, we do not consider the magnitude of the difference. Therefore, we use ReLU function (written as $\sigma(\cdot)$ ) to penalize positive $\left(V_{\phi}\left(\delta^*,\omega^*)\right)-V_{\phi}(\delta,\omega)\right )$.}  Define the loss term as:
    \begin{equation}\label{eq:Loss_Lya_2}
        l_2(\phi)=\frac{1}{H}\sum_{h=1}^H \sigma\left(-V_{\phi}(\delta^h,\omega^h)+V_{\phi}\left(\delta^*,\omega^*)\right)\right )
    \end{equation}
    
    \item \textit{Penalty term with $\nabla_{f_{u}} V_{\phi}(\delta^*,\omega^*)$}
    
    This term is employed to mitigate numerical errors.
    We design a extra loss term to penalize on the value of $\nabla_{f_{u}} V_{\phi}(\delta^*,\omega^*)$ as:
    \begin{equation}\label{eq:Loss_Lya_3}
        l_3(\phi)=\left(\nabla_{f_{u}} V_{\phi}(\delta^*,\omega^*)\right)^2+\sigma\left(\nabla_{f_{u}} V_{\phi}(\delta^*,\omega^*)\right)
    \end{equation}
    where $\left(\nabla_{f_{u}} V_{\phi}(\delta^*,\omega^*)\right)^2$ guarantee the small magitute of $\nabla_{f_{u}} V_{\phi}(\delta^*,\omega^*)$. Considering that $\nabla_{f_{u}} V_{\phi}(\delta^h,\omega^h)$ should never be positive, 
    \highlight{we use ReLU function $\sigma\left(\nabla_{f_{u}} V_{\phi}(\delta^*,\omega^*)\right)$ to guarantee that $\nabla_{f_{u}} V_{\phi}(\delta^*,\omega^*)$ is negative close zero. }
     This way, the zero action at the equilibrium is guaranteed to satisfy Lyapunov conditions.
\end{enumerate}


Combining~\eqref{eq:Loss_Lya_1}-\eqref{eq:Loss_Lya_3}, the total loss function is 
    \begin{equation}\label{eq:Loss_total}
       L_{q}(\boldsymbol{\phi}) =q_1  l_1(\phi)+q_2  l_2(\phi)+q_3  l_3(\phi)
    \end{equation}
where $q_1$, $q_2$, $q_3$ are hyperparameters balancing the loss terms, with $q_3$ tuned to be much larger than the others. \highlight{ For the specific problem in this manuscript, we found that letting magnitude of $q_1$ to be slight larger than $q_2$ (e.g., $q_1$ to be 1.5 times of $q_2$) leads to most samples satisfy Lyapunov conditions. }Note that the equilibrium $(\delta^*,\omega^*)$ is obtained from the steady state in \eqref{eq:Dynamic} and we fix the equilibrium in training. Of course the equilibrium changes if the load or the parameters changes. More specifically, $\omega^*=0$ always while $\delta^*$ varies. Since we only use the learned function as a regularization to train a controller, we are robust to changes in the equilibrium point. If the learned function is used to certify stability, then the changes in equilibrium should be carefully accounted for.

\subsection{Algorithm with Active Sampling}

The goal for training the neural Lyapunov function is to make larger proportional of the batch samples satisfy the conditions~\eqref{eq:Lyapunov_Condition}. The pseudo-code for our proposed method is given in Algorithm 1. A linear controller is used to initialize training. Let $\varrho$ be the proportion of samples that satisfy the conditions~\eqref{eq:Lyapunov_Condition}.
After most of the samples (e.g., $\varrho>95\%$) have already satisfied the conditions, it would be difficult to improve the neural Lyapunov function further since the loss function will remain almost unchanged even though $\varrho$ increases slightly. We augment the training performance by collecting samples violate~\eqref{eq:Lyapunov_Condition} and add them to the next batch of training. \highlight{Moreover, since we care more about the region with smaller frequency deviation, we also let 50\% of the batch states to be sampled from regions close to equilibrium.} This way, the neural Lyapunov can improve efficiently and $\varrho$ can reach 99.9\% in the end \highlight{for both the region close and away from the equilibrium}. Adam algorithm is adopted to update weights $\bm{\phi}$ in each episode.


\begin{algorithm}
 \caption{Learning neural Lyapunov function}
 \begin{algorithmic}[1]
 \renewcommand{\algorithmicrequire}{\textbf{Require: }}
 \renewcommand{\algorithmicensure}{\textbf{Input:}}
 \REQUIRE  Learning rate $\alpha$, number of episodes $I$, state transfer function~\eqref{eq:Dynamic}, hyperparameters in~\eqref{eq:Loss_Lya_1}-\eqref{eq:Loss_total}
 \ENSURE  Droop coefficient $l_i$  for the $i$-th bus, $i=1,\cdots,N$ \\
\textit{Initialisation} :Initial weights $\phi$ for neural network
  \FOR {$episode = 1$ to $I$}
  \STATE Generate batch state samples $\delta^h,\omega^h$ for the $h$-th batch, $h=1,\cdots,H$\\
  \STATE If $\varrho>\bar{\varrho}$, add the samples violates Lyapunov condition
  $\{(\delta,\omega)\}\leftarrow \{(\delta,\omega),(\hat{\delta},\hat{\omega} )\}$\\
  \STATE Compute $f_u(\delta,\omega)$ for the sample states with linear droop control using  \eqref{eq:Dynamic}\\
  \STATE Calculate $V_{\phi}\left(\delta,\omega\right)$ and $\nabla_{f_{u}} V_{\phi}\left(\delta,\omega\right)$\\
  \STATE Identify the states $\left ( \hat{\delta},\hat{\omega} \right )$ that does not satisfy Lyapunov condition and its percentage  $\varrho$ 
  \\
  \STATE Calculate total loss of all the batches using \eqref{eq:Loss_Lya_1}-\eqref{eq:Loss_total}\\
  \STATE Update weights in the neural network by passing $Loss$ to Adam optimizer:
  $\bm{\phi} \leftarrow \bm{\phi}-\alpha \text{Adam}(Loss)$
  \ENDFOR
 \end{algorithmic} 
 \end{algorithm}

%% file: Controller.tex
 We propose to use the learned neural Lyapunov function to guide the training of neural network controller. We adopt the neural Lyapunov function as an additional regularization that is used during the training process of the neural network controller. The real-time control policy is computed through the \highlight{feedforward neural networks where the input is the local frequency deviation and the weights are trained offline.}
 Note that we may be able to achieve better performance through a projection if the Lyapunov conditions are violated. However, such a projection requires information of all the state variables in real-time, which is unrealistic for the power system with large numbers of nodes and limited communication.

 \subsection{Lyapunov Regularization} 
 Given a Lyapunov function, Proposition~\ref{prop:exponential stable} illustrates the condition for locally exponentially stability~\cite{aastrom2010feedback}.
 
 \begin{prop}[locally exponentially stable condition]\label{prop:exponential stable}
For the function $V: \mathcal{D} \rightarrow \mathbb{R}$ satisfying \eqref{eq:Lyapunov_Condition}, if there is constant $\beta >0$ such that for all  $(\delta,\omega)\in\mathcal{D}$ we have
\begin{equation}\label{eq:Lya_exponential}
  \nabla_{f_{u}} V(\delta,\omega) \leq-\beta  \left(V(\delta,\omega)-V(\delta^*,\omega^*) \right)  
\end{equation}
Then, the equilibrium is locally exponentially stable.
 \end{prop}
 
In order to satisfy~\eqref{eq:Lya_exponential} with the neural network controller, we propose a Lyapunov regularization approach that the action is penalized if this inequality does not hold. Compared with traditional regularization (e.g., lasso, ridge) or penalty term on large state magnitude, we do not add regularization uniformly to all the weights or actions. Instead, the action is only penalized when~\eqref{eq:Lya_exponential} is violated. The regularization term is
$$R_{\phi}(u_{\theta})=\sigma\left(\nabla_{f_{u}} V_{\phi}(\delta,\omega)+\beta  (V_{\phi}(\delta,\omega)-V_{\phi}(\delta^*,\omega^*))\right)
$$


\subsection{Controller and Architecture }
The formulation of controller and the training architecture is from our previous work~\cite{cui2020}. For completeness, we reiterate the key design in this subsection. The work in~\cite{cui2020} showed that a controller mapping frequency to active power needs to be a function that is monotonic, increasing and goes through the origin. To this end, we explicitly engineer the neural network controller with a stacked-ReLU structure and represented as~\eqref{eq:stacked-relu}
\begin{subequations}\label{eq:stacked-relu}
\begin{align}
     &u_i(\omega_i)=\color{black} s_i\sigma(\mathbf{1}\omega_i+b_i)+z_i\sigma(-\mathbf{1}\omega_i+c_i)\\
     \mbox{where}
     &\sum_{j=1}^{l} s_i^{j}\geq 0, \quad \sum_{j=1}^{l} z_i^{j}\leq 0, \quad \forall l=1,2,\cdots,m
     \label{eq:f+2}\\
     & b_i^{1}=0, b_i^{l}\leq b_i^{(l-1)},\quad \forall l=2,3,\cdots,m\\
     &c_i^{1}=0, c_i^{l}\leq c_i^{(l-1)},\quad \forall l=2,3,\cdots,m\label{eq:f+3}
\end{align}
\end{subequations}
where $m$ is the number of hidden units and $\mathbf{1}\in \mathbb{R}^m$ is the all $1$'s column vector. Variables $\color{black}s_i=[\begin{matrix}
 s_i^{1}& s_i^{2} & \cdots&s_i^{m}
\end{matrix}]$ and $z_i=[\begin{matrix} z_i^{1}&z_i^{2}&\cdots&z_i^{m}\end{matrix}]$ are the weight vector of bus $i$; $b_i=[\begin{matrix} b_i^{1}& b_i^{2} & \cdots &b_i^{m}\end{matrix}]^\intercal$ and $c_i=[\begin{matrix} c_i^{1}& c_i^{2} & \cdots & c_i^{m}\end{matrix}]^\intercal$ are the corresponding bias vector. The variables to be trained are weights $\bm{\theta}=\{s,b,z,c\}$ in \eqref{eq:stacked-relu}.

To obtain the trajectory for training the controller, we discretize dynamics \eqref{eq:Dynamic} with step size $\Delta t$. We use $k$ and $K$ to represent the discrete time and total number of stages, respectively. The neural network controller is then denoted as $u_{\theta_i}(\omega_i)$. From \eqref{eq:Dynamic}, $\omega_i(k)$ in each timestep $k$ is a function of  $\omega_i(k-1)$ and $u_{\theta_i}(\omega_i(k-1))$, which is then a function of $\omega_i(k-2)$ and $u_{\theta_i}(\omega_i(k-2)$. This means that computing gradient of $u_{\theta_i}(\omega_i(k)$ with respect to $\theta_i$ needs the chain-rule from the step k all the way to the first time step for all $k=0, \cdots, K$. To mitigate the computation burden caused by the subsequent application of chain-rule, we proposed a RNN-based framework to integrate the state transition dynamics~\eqref{eq:Dynamic} implicitly.

\begin{figure}[H]	
	\centering
	\includegraphics[width=3.38in]{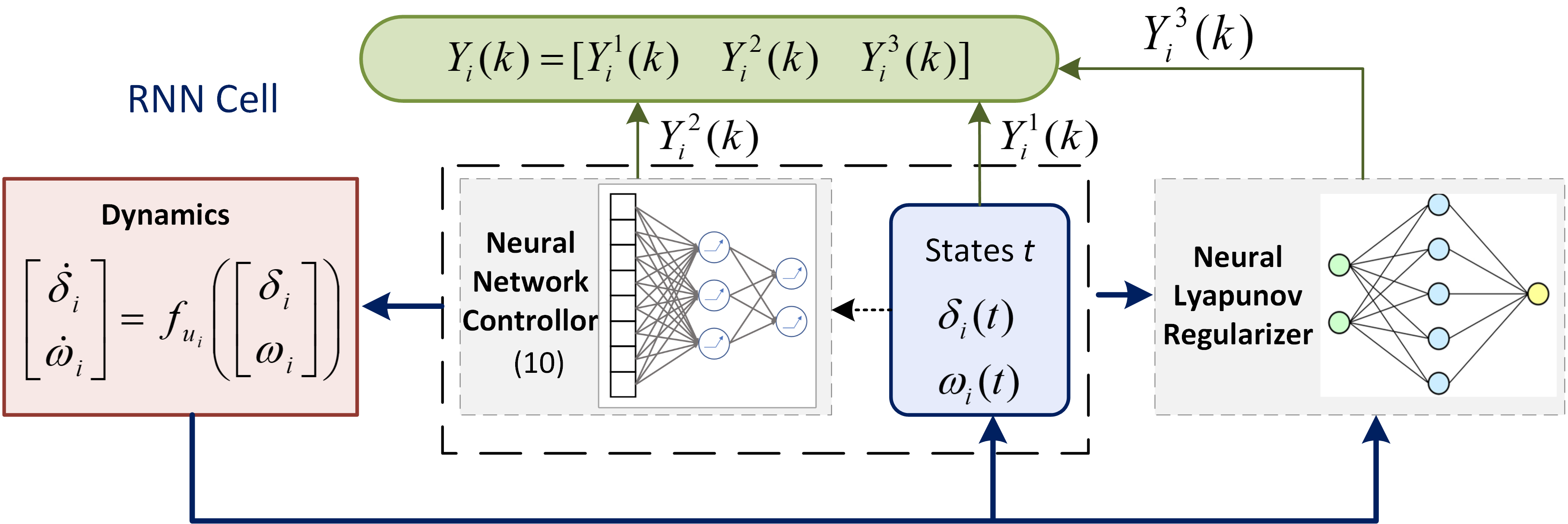}
	\caption{Structure of RNN for frequency control problem}
	\label{fig:RNN_Structure}
\end{figure}

As illustrated in Fig.\ref{fig:RNN_Structure}, the state of RNN cell of bus $i$ is set to be $\left(\delta_i,\omega_i\right)$. The system dynamics \eqref{eq:Dynamic} is set as the transition function of RNN cell.
At each time $k$, state of RNN cell and the current action from neural network controller will go through the transition dynamics to calculate the state of the time $k+1$. The state $\omega$ and action $u$ constitute the first two component of output where $Y_i^1(k)=\omega_i(k)$ and $Y_i^2(k)=\left (u_{\theta_i}(\omega_i(k))  \right )^2$.
The total state information and time derivative information are simultaneously send as input into Neural Lyapunov function to calculate the Lyapunov regularization term, written as $Y_i^3(k)=\sigma\left(\nabla_{f_{u}} V_{\phi}(\delta,\omega)+\beta  (V_{\phi}(\delta,\omega)-V_{\phi}(\delta^*,\omega^*))\right)/N$. 

The loss function is formulated to be equivalent with the objective function \eqref{subeq:Continuous_Optimization_obj} plus the Lyapunov regularization as:
\begin{equation}\label{eq:RNN_Loss}
\begin{split}
    Loss=&\sum_{i=1}^{N} \max_{k=0,\cdots,K} |Y_i^1(k)|+\gamma\frac{1}{K}\sum_{k=1}^{K}Y_i^2(k)\\
    &+\lambda\frac{1}{K}\sum_{k=1}^{K} Y_i^3(k). 
\end{split}
\end{equation}



    

\subsection{Algorithm to Train Neural Network Controller}
The pseudo-code for learning the neural network controller is given in Algorithm 2.  Training is implemented in a batch updating style where the $h$-th batch initialized with randomly generated initial states $\{\delta_i^h(0),\omega_i^h(0)\}$ for all $i=1,\cdots,N$. The evolution of states in $K$ stages will be computed through structure of RNN as shown by Fig.\ref{fig:RNN_Structure}.  Although algorithms 1 and 2 can be iterated to make further update, we did not see an obvious improvement in simulation.  

\begin{algorithm}
 \caption{Reinforcement Learning with RNN}
 \begin{algorithmic}[1]
 \renewcommand{\algorithmicrequire}{\textbf{Require: }}
 \renewcommand{\algorithmicensure}{\textbf{Input:}}
 \REQUIRE  Learning rate $\alpha$, batch size $H$, total time stages K, number of episodes $I$, parameters in optimal frequency control problem \eqref{eq:Continuous_Optimization}
 \ENSURE  The neural Lyapunov function $V_{\phi}(\delta,\omega)$\\
\textit{Initialisation} :Initial weights $\theta$ for control network
  \FOR {$episode = 1$ to $I$}
  \STATE Generate initial states $\delta_i^h(0),\omega_i^h(0)$ for the $i$-th bus in the $h$-th batch, $i=1,\cdots,N$, $h=1,\cdots,H$\\
  \STATE Reset the state of cells in each batch as the initial value $x_i^h\leftarrow \{\delta_i^h(0),\omega_i^h(0)\}$.\\
  \STATE RNN cells compute through K stages to obtain output $\{Y_{h,i}(0),Y_{h,i}(1),\cdots,Y_{h,i}(K)\}$ \\
  \STATE Calculate total loss of all the batches $Loss=\frac{1}{H}\sum_{h=1}^{H} \sum_{i=1}^{N} \max_{k=0,\cdots,K} |Y_{h,i}^1(k)|+\gamma\frac{1}{K}\sum_{k=1}^{K}Y_{h,i}^2(k)+\lambda\frac{1}{K}\sum_{k=1}^{K} Y_{h,i}^3(k)$.\\
  \STATE Update weights in the neural network by passing $Loss$ to Adam optimizer:
  $\bm{\theta} \leftarrow \bm{\theta}-\alpha \text{Adam}(Loss)$
  \ENDFOR
 \end{algorithmic} 
 \end{algorithm}

%% file: simulation.tex
Case studies are conducted on the IEEE New
England 10-machine 39-bus (NE39) power network\highlight{~\cite{chow1992toolbox}} to illustrate the effectiveness of the proposed method. We visualized the learned Lyapunov function and its Lie derivative. Then we show that regularization is necessary, in the sense that a controller learned without it can be unstable. Lastly, we show the training losses. 
\subsection{Simulation Setting}\label{subsec:Simulation Setting}
 The step size for the discrete simulation is set to 0.02 (20ms) and the time stages K is 100. \highlight{Power injection $P_{i}$ }are set at the nominal values, the bound on action $\overline{u}_i$ is uniformly distributed in $[0.8P_{i}, P_{i}]$ and $\gamma$ is set as 0.005. \highlight{The parameters for training the neural networks are given in Appendix~\ref{app:parameters}.}

\subsection{Visualization Lyapunov Function and the Lie Derivative }
\begin{figure}[ht]	
	\centering
	\includegraphics[width=3.36in]{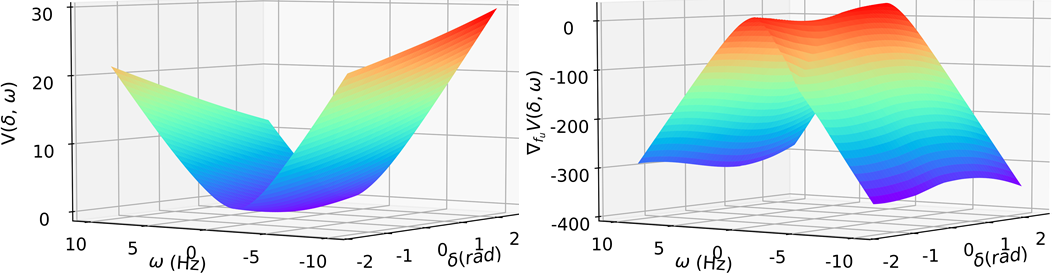}
	\caption{Neural Lyapunov function (left) and Lie derivative (right) when changing $(\delta,\omega)$ in generator 5 and keep state variable of other generators at the equlibrium value.}
	\label{fig:Lyapunov_plot}
\end{figure}

To visualize the Lyapunov function with a large number of state variables, we fix all the states at their equilibrium value and vary the state variable for one generator bus. Fig.~\ref{fig:Lyapunov_plot} illustrates the value of Lyapunov function and Lie derivative with the variation of $\delta$ and $\omega$ in generator bus 5. The Lyapunov function $V(\delta,\omega)$ achieves the minimum at the equilibrium point and thus satisfies condition  \eqref{eq:Lyapunov_Condition_V}. The Lie derivative $\nabla_{f_{u}} V_{\phi}\left(\delta,\omega\right)$ is smaller than zero in most of the regions and thus also generally satisfy condition \eqref{eq:Lyapunov_Condition_dot_V}. After convergence, only 0.1\% of samples with $\omega$ sufficiently close to zero make $\nabla_{f_{u}} V_{\phi}\left(\delta,\omega\right)$ to be slightly positive. Such a small positive number only leads to small Lyapunov regularization term and therefore has neglectable impact on the training of neural network controller. 



\subsection{Performance Comparison}
Under the same hyperparameters and RNN structure, we train the neural network controller with Lyapunov regularization (labeled as RNN-Lyapunov) and without Lyapunov regularization (labeled as RNN-w.o.-Lyapunov), respectively. \highlight{We test the effect of large deviation in initial operating points and sudden changes in topology.}

\begin{figure}[ht]
        \centering
        \subfloat[Dynamics of $u$ (left) and $\omega$ (right) for RNN with Lyapunov regularization  ]{\includegraphics[width=3.38in]{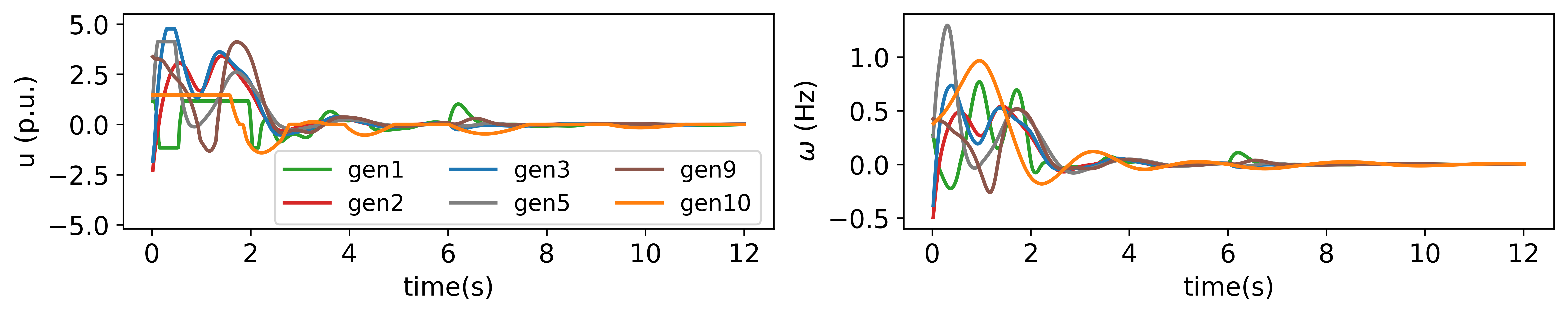}}
        \hfil
        \subfloat[Dynamics of $u$ (left) and $\omega$ (right) for linear droop control ]{\includegraphics[width=3.38in]{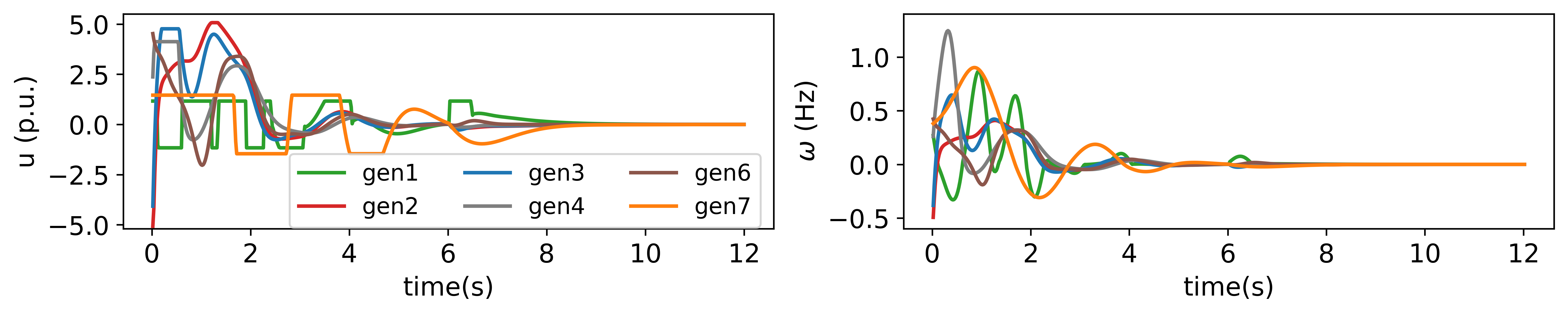}%
        }
        \hfil
        \subfloat[Dynamics of $u$ (left) and $\omega$ (right) for RNN without Lyapunov regularization ]{\includegraphics[width=3.38in]{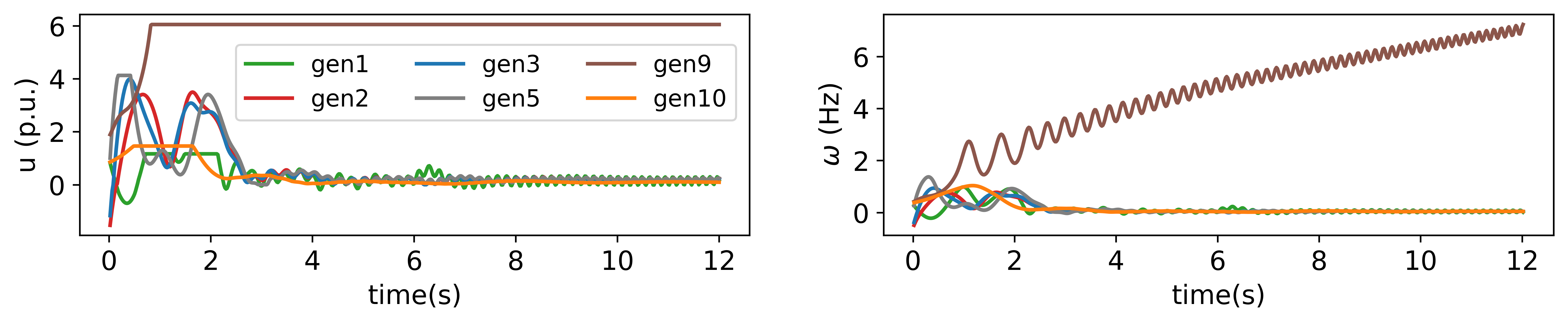}%
        }
        \caption{Dynamics of control action $u$ and frequency deviation $w$ in selected generator buses corresponding to (a) RNN-Lyapunov (b) linear droop (c) RNN-w.o.-Lyapunov. The neural network controller trained with Lyapunov regularization achieve smaller control cost than linear droop control, and better stablizing performance than that without Lyapunov regularization.}
        \label{fig:Dynamic_Behavior_r1}
        \end{figure}

\highlight{At time t=0, the system starts from some initial conditions that deviates from the equilibrium. At time t=6s, the lines between buses 1 and 39, and 2 and 3 are disconnected. The dynamics of the system under the controller obtained by RNN-Lyapunov, linear droop control and RNN-w.o.-Lyapunov are shown in Fig.~\ref{fig:Dynamic_Behavior_r1}. Both RNN-Lyapunov and linear droop control stablize the system, while RNN-w.o.-Lyapunov leads to divergence as shown in Fig.~\ref{fig:Dynamic_Behavior_r1}(c). Compared with dynamics of linear droop control in Fig.~\ref{fig:Dynamic_Behavior_r1}(b), RNN-Lyapunov in Fig.~\ref{fig:Dynamic_Behavior_r1}(a) achieve similar frequency deviation while using smaller control action.}

\highlight{After losing two lines at time t=6s, the system experience frequency deviation of approximate 0.1 Hz and return to stable state within 2s for RNN-Lyapunov (Fig.~\ref{fig:Dynamic_Behavior_r1}(a)). Therefore, the proposed RNN-Lyapunov approach is robust to topology changes. In Appendix~\ref{app:warm start}, we also show that the existing weights for the old topology can serve as good initialization for training with the new topology.}

\begin{figure}[ht]	
	\centering
	\includegraphics[width=2.8in]{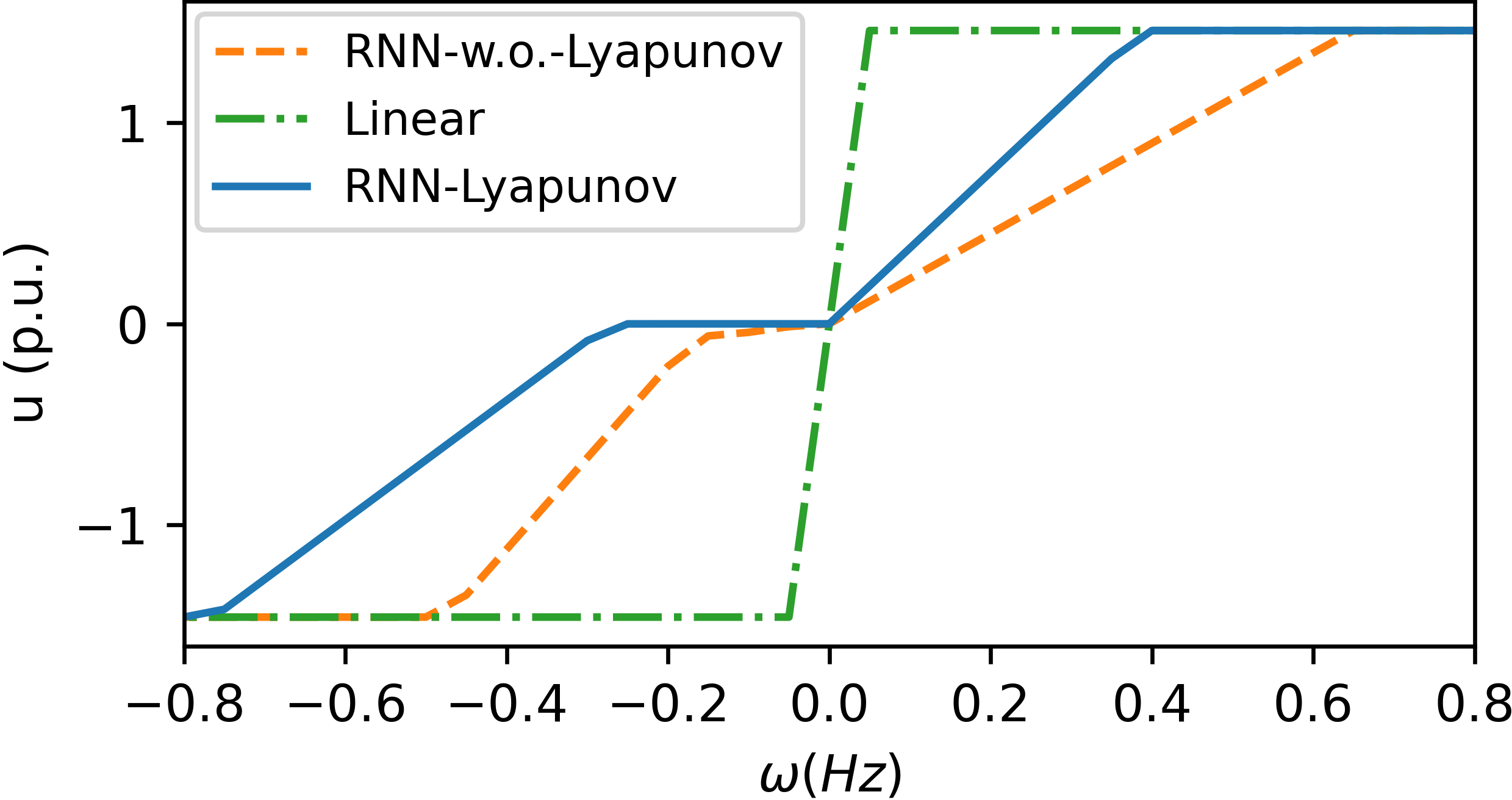}
	\caption{Control action $u$ of RNN-Lyapunov, RNN-w.o.-Lyapunov and Linear droop control for generator bus 10. Lyapunov regularization leads to different non-linear control law.}
	\label{fig:Action}
\end{figure}

\begin{figure}[ht]	
	\centering
	\includegraphics[width=2.8in]{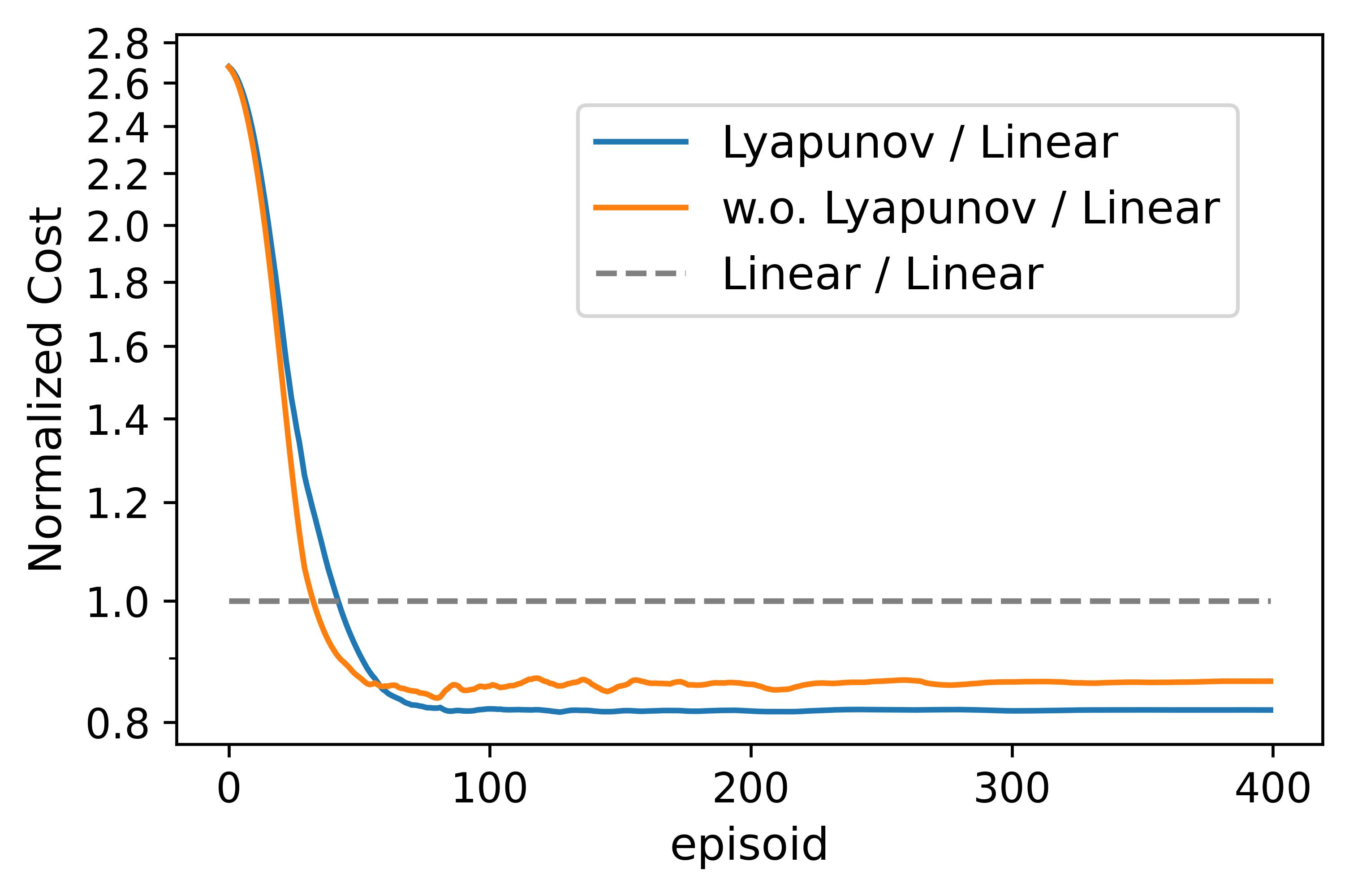}
	\caption{Normalized cost along the episode during the training of neural network controller with and without Lyapunov regularization. RNN-Lyapunov and RNN-w.o.-Lyapunov reduce the cost by approximate 19\% and 14\% compared with linear droop control.}
	\label{fig:Loss_plot}
\end{figure}

We further compare RNN-Lyapunov and RNN-w.o.-Lyapunov with the benchmark of linear droop control, where the droop coefficient is obtained by solving problem~\eqref{eq:Continuous_Optimization} using fmincon function of Matlab~\cite{cui2020}. Fig.~\ref{fig:Action} illustrates the control policy obtained from the three methods.
Compared with linear droop control, the stacked-ReLU neural network learns a highly non-linear controller. The average cost normalized by the cost of linear droop control along episode is shown in Fig.~\ref{fig:Loss_plot}. Both RNN Lyapunov and RNN-w.o.-Lyapunov converge in approximate 150 episodes. After convergence, RNN-Lyapunov reduces the cost by approximate 19\% compared with linear droop control. The reduction is 5\% more than that of RNN-w.o.-Lyapunov. Therefore, the proposed method learns a non-linear stabilizing controller that performs better than traditional linear droop control.
\highlight{Additional numerical validation with power disturbance and larger test system can be found in  Appendix~\ref{app:power disturbance} and Appendix~\ref{app:300-bus}, respectively. Computational time is also provided in Appendix~\ref{app:300-bus}.}

%% file: appendix.tex
\section{Simulation Parameters}\label{app:parameters}
 The parameters for training the neural networks are:
\begin{itemize}
    \item Neural Lyapunov function is parameterized as a dense neural network with one hidden layer of 50 neurons and ELU activation. The episode number is 4000. The hyper-parameters in \eqref{eq:Loss_Lya_1}-\eqref{eq:RNN_Loss} are $\mu=50$, $q_1=10$, $q_2=5$, $\color{black}q_3=100$, $\beta=0.005$, $\lambda=0.01$. \highlight{ Each episode has the batch number of 500 with random states samples. For 50\% of the batch samples, $\delta_i^h$ is uniformly distributed in $[-2,2]\,\text{rad}$, $\omega_i^h$ is uniformly distributed in $[-10,10]\,\text{Hz}$. For another  50\% of the batch samples, states are sampled from normal distribution with its mean at the equilibrium and truncated after 2 standard deviations. The stand deviation for $\delta$ and $\omega$ is set to be 0.2 and 0.5, respectively.} Trainable weights are updated using Adam with learning rate initializes at 0.05 and decay every 100 steps with a base of 0.9.

    

    \item Neural network controller is parameterized as the stacked-ReLU function \eqref{eq:stacked-relu} with 20 neurons ($m=20$). The episode number is 400 \highlight{ and the batch number is 100}. \highlight{To simulate the random initial states of a post-fault system~\cite{chiang1989study, arapostathis1982global}, we let $\delta_i(0)$ be uniformly distributed in $[-1,1]\,\text{rad}$ and $\omega_i(0)$ be uniformly distributed in $[-0.5,0.5]\,\text{Hz}$. }
    Trainable weights are updated using Adam with learning rate initializes at 0.04 and decay every 30 steps with a base of 0.7.
    
\end{itemize}

\section{Simulation for Step Changes}\label{app:power disturbance}
\begin{figure}[ht]
\subfloat[Dynamics of $u$ (left) and $\omega$ (right) for RNN-Lyapunov ]{\includegraphics[width=3.38in]{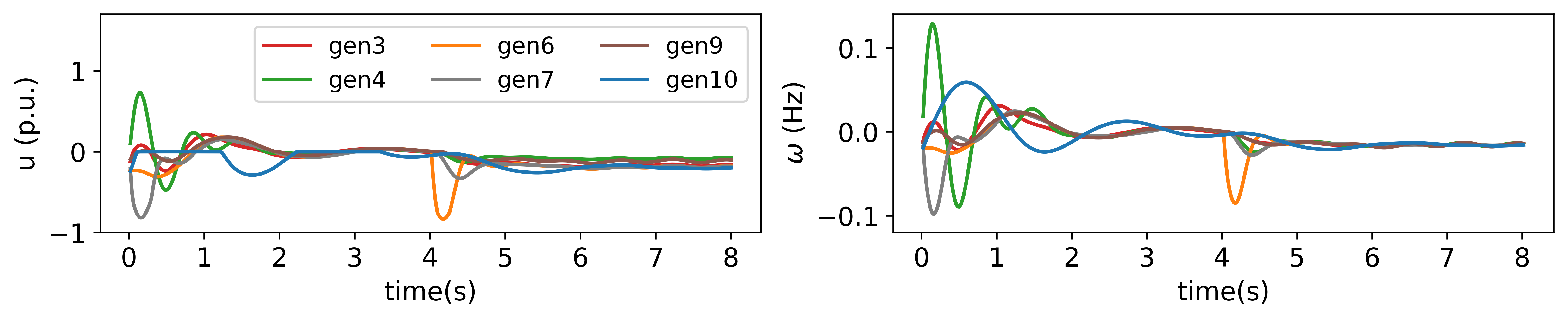}%
\label{fig_first}}
\hfil
\subfloat[Dynamics of $u$ (left) and $\omega$ (right) for linear droop control ]{\includegraphics[width=3.38in]{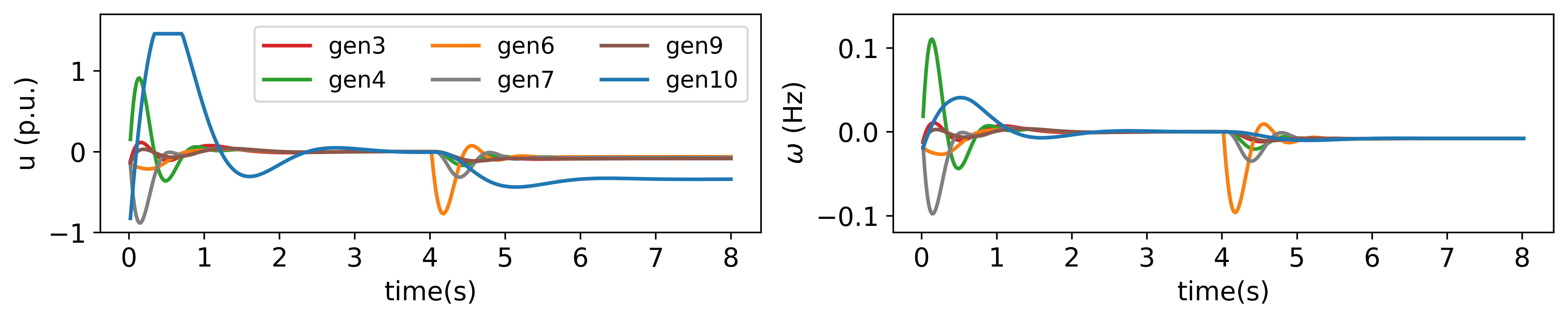}%
}
\label{fig_second}
\hfil
\subfloat[Dynamics of $u$ (left) and $\omega$ (right) for RNN-w.o.-Lyapunov ]{\includegraphics[width=3.38in]{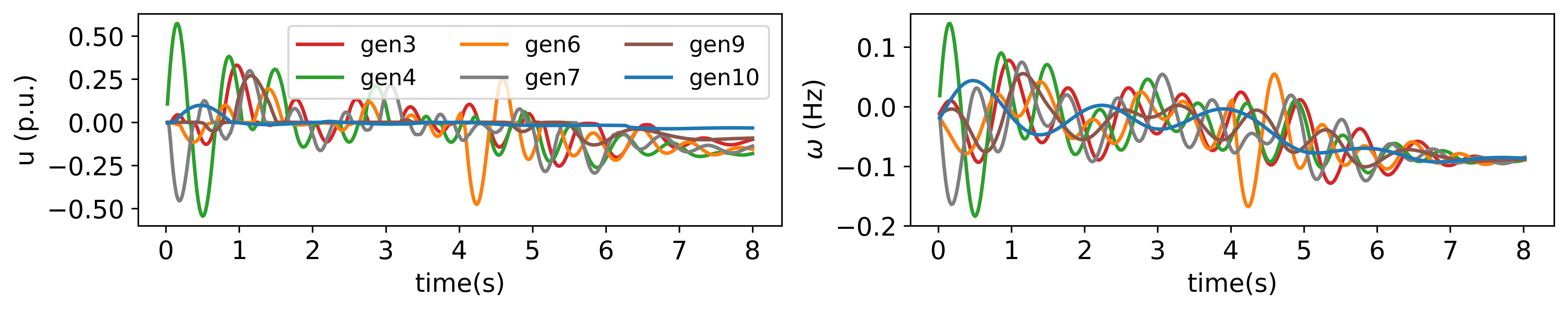}%
\label{fig_third}}
\caption{Dynamics of control action $u$ and frequency deviation $w$ in selected generator buses corresponding to (a) RNN-Lyapunov (b) linear droop (c) RNN-w.o.-Lyapunov . The neural network controller trained with Lyapunov regularization achieve smaller control cost than linear droop control, and better stablizing performance than that without Lyapunov regularization.}
\label{fig:Dynamic_Behavior}
\end{figure}

\highlight{Fig.\ref{fig:Dynamic_Behavior} are the dynamics of $\omega$  and corresponding control action $u$ for different controllers with the same initial condition and the loss of 50\% of generation capacity in generator 6 after t=4s.  Compared with linear droop control in Fig.\ref{fig:Dynamic_Behavior}(b), RNN-Lyapunov in Fig.\ref{fig:Dynamic_Behavior}(a) achieves similar frequency performance with much smaller control efforts. For controllers based on RNN-w.o.-Lyapunov in  Fig.\ref{fig:Dynamic_Behavior}(c), the states do not converge well and continue to oscillate at the end of the simulation.} 


\section{Warm Start for Topology Changes}\label{app:warm start}
\begin{figure}[ht]	
	\centering
	\includegraphics[width=2.8in]{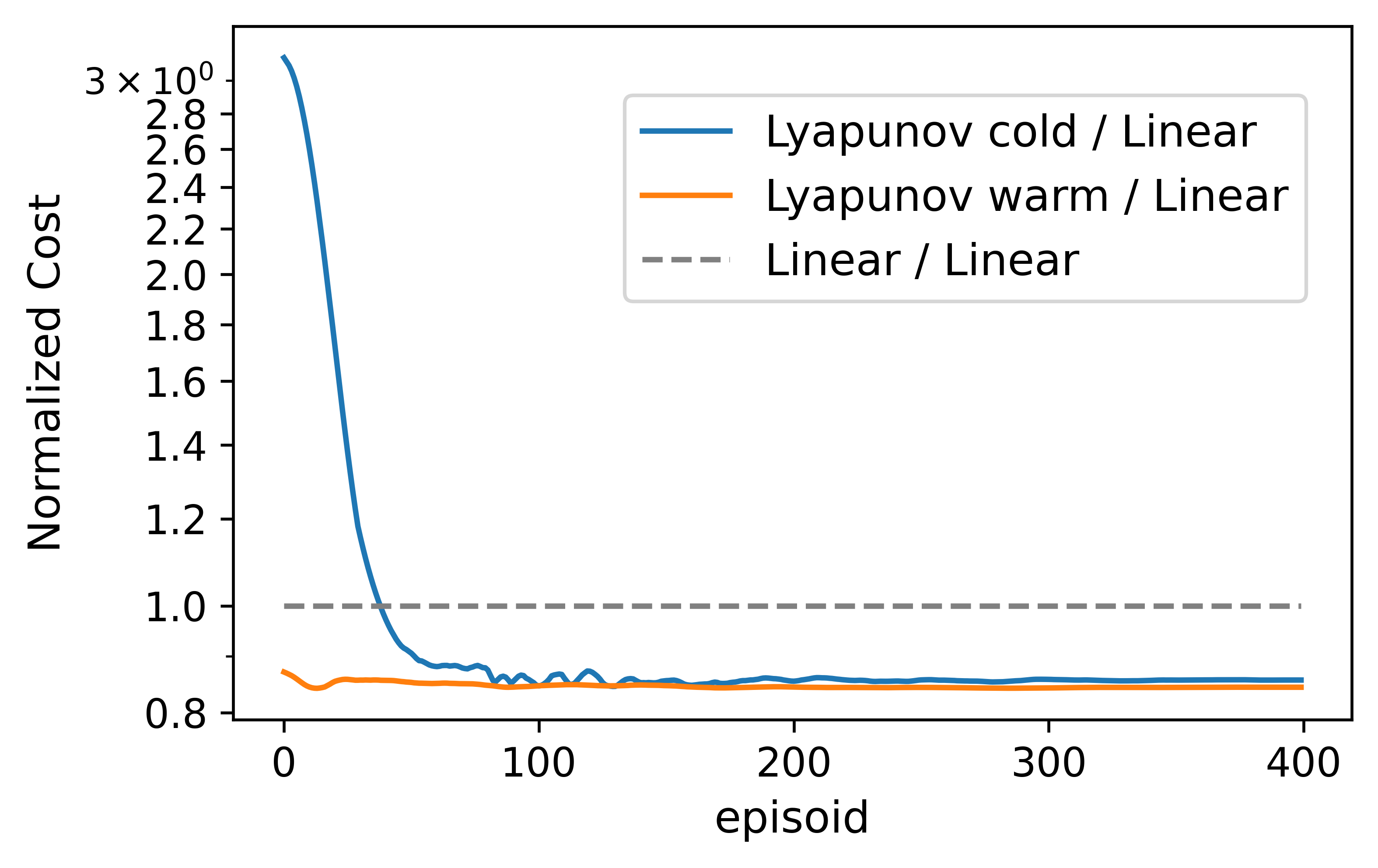}
	\caption{Normalized cost along the episode during the training of neural network controller with and without warm start from existing weights.  Compared with Lyapunov cold that converges after approximate 200 episodes, Lyapunov warm converges in approximate 20 episodes and therefore speeds up the training process tremendously.}
	\label{fig:warm_start}
    \end{figure}

   \highlight{All of the previous results are reported for a controller trained from scratch. Of course, we need not to restart the entire process if the topology of the grid changes. The existing weights for the old topology can serve as good initialization for training the new controller.
    Fig.~\ref{fig:warm_start} illustrates the average cost normalized by the cost of linear droop control along episode for training with warm start (referred to as Lyapunov warm) and without warm start (referred to as Lyapunov cold). Compared with Lyapunov cold that converges after approximate 200 episodes, warm starts converges in approximate 20 episodes and therefore speeds up the training process significantly. 
    }

\section{Test on IEEE 300-Bus Test System}\label{app:300-bus}
\highlight{In this subsection, we further conduct a case study on the IEEE 300-bus test system~\cite{chow1992toolbox} to validate the performance of our proposed method in a larger system. We modify the system such that there are 100 generators in the network (the original network has 69 generators) to create a moderately large test case.  
The parameters for the test system are set according to~\cite{nishikawa2015comparative}. Batch size and the episode number for training the Lyapunov function is 1000 and 6000, respectively.  Batch size and the episode number for training the neural network controller is 150 and 400, respectively. Other parameter setting for the neural network controller and the training process are the same as the 39-bus system. }

\begin{figure}[ht]	
	\centering
	\includegraphics[width=2.8in]{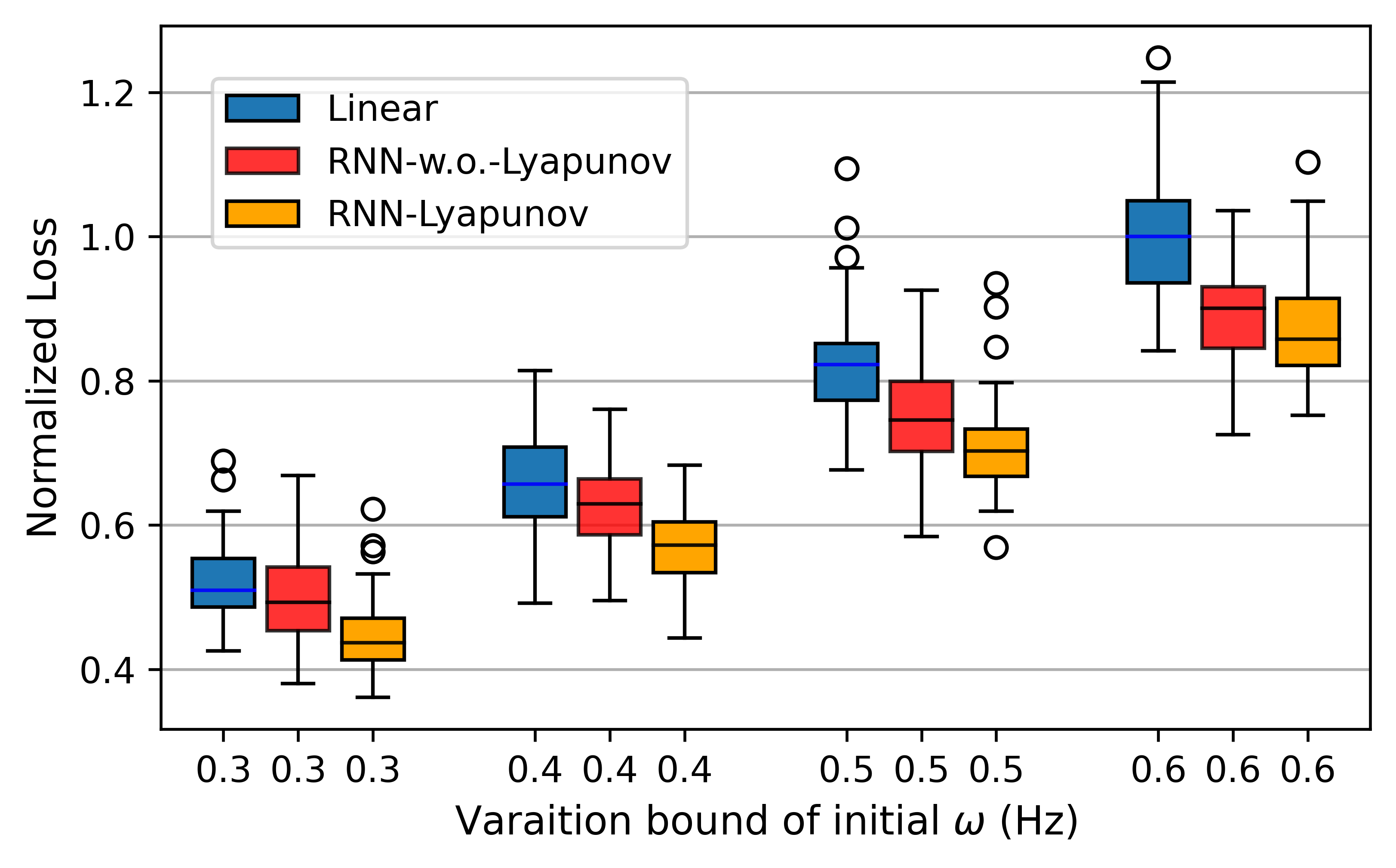}
	\caption{Loss with different variation range of initial conditions for RNN-Lyapunov, RNN-w.o.-Lyapunov and Linear droop controller. RNN-Lyapunov achieves smallest loss.  }
	\label{fig:Loss_100gen}
\end{figure}

\highlight{To investigate the general performance of  RNN-Lyapunov and linear droop with different initial conditions, we fix the distribution of initial $\delta$ and let the initial $\omega$ to uniformly distributed in $[-\bar{\omega},\bar{\omega}]$.  The distribution of loss corresponding to $\bar{\omega} =0.3,\cdots,0.6\,\text{Hz}$ are illustrated as box plot in Fig.~\ref{fig:Loss_100gen}. We normalize the loss with the maximum median value corresponding to linear droop for ease of comparison. For  $\bar{\omega}=0.6$, the median of RNN Lyapunov and RNN-w.o.-Lyapunov is 0.8573, which is approximate 14.27\% lower than that of linear droop control (=1) and 4.31\% lower than RNN-w.o.-Lyapunov (=0.9004). This trend also holds in general with the decrease of $\bar{\omega}$.  Therefore, the non-linear controller obtained by RL achieves lower cost than the traditional linear droop control. Lyapunov regularization further reduce the cost corresponding to neural network controller by reducing high cost of diverged trajectory. 
}
        
\highlight{We use TensorFlow 2.0 framework to build the reinforcement learning environment and run the training process in Google Colab with GPU acceleration. For the 300-bus test system, the time for training the Lyapunov function and training the neural network controller is 209s and 872s, respectively. These times only increase moderately compared to the 39-bus test system, which used 117s for training Lyapunov function and 735s for training the neural network controller, respectively. }